\documentclass[%
 reprint, nofootinbib,
 amsmath,amssymb,
 aip,
]{revtex4-1}
\usepackage[english]{babel}
\usepackage{graphicx}
\usepackage{dcolumn}
\usepackage{textcomp}
\usepackage{array}
\usepackage{bm}
\usepackage{xcolor}
\usepackage{amsmath, esint}
\usepackage[normalem]{ulem}
\newcommand{\ve}[1]{\mathbf{#1}}
\newcommand{\fig}[1]{Figure~\ref{#1}}
\newcommand{\eq}[1]{Eq.~(\ref{#1})}
\newcommand{\eqs}[2]{Eqs.~(\ref{#1}) and (\ref{#2})}
\newcommand{\tab}[1]{Table~\ref{#1}}
\newcommand{\corr}[1]{#1}
\newcommand{\correc}[1]{#1}

\begin{document}
\preprint{AIP/123-QED}
\title{Instability-enhanced transport in low temperature magnetized plasma}
\author{R. Lucken}
  \email{romain.lucken@lpp.polytechnique.fr}
\author{A. Bourdon}
\affiliation{Laboratoire de Physique des Plasmas, CNRS, Sorbonne Universit\'e, Universit\'e Paris Sud, \'Ecole Polytechnique, F-91120 Palaiseau, France}
\author{M. A. Lieberman}%
\affiliation{Department of Electrical Engineering and Computer Sciences, University of California, Berkeley, California 94720 }
\author{P. Chabert}
\affiliation{Laboratoire de Physique des Plasmas, CNRS, Sorbonne Universit\'e, Universit\'e Paris Sud, \'Ecole Polytechnique, F-91120 Palaiseau, France}
 \homepage{http://www.lpp.fr}
\date{\today}
\begin{abstract}
It is shown that the transport in low temperature, collisional, bounded plasma is enhanced by instabilities at high magnetic field. 
While the magnetic field confines the electrons in a stable plasma, the instability completely destroys the confinement such that the transport becomes independent of the magnetic field in the highly magnetized limit. 
\correc{An analytical expression of the instability-enhanced collision frequency is proposed, based on a magnetic field independent edge-to-center density ratio. }

\end{abstract}
\maketitle
In low temperature plasma discharges, the transport from the ionization region to the walls determines the electron temperature and hence the global discharge properties. 
The classical drift-diffusion theory of low density magnetized plasma transport \cite{simon,fruchtman,sternberg} predicts that the plasma should reach a steady-state equilibrium qualitatively analogous to that of a non-magnetized discharge \cite{tonks,schottky,godyak} and that the confinement increases with the magnetic field. However, when the value of the magnetic field is high enough, strong instabilities develop that deconfine the electrons, which enhance the macroscopic transport \cite{curreli}. These so-called universal instabilities develop in bounded plasma, without any external source of energy. The wave energy comes from the natural drift-diffusion motion of the particles.\cite{chen} They can be either Simon-Hoh instabilities (similar to a Rayleigh-Taylor instability with an electrostatic potential),\cite{boeuf} or electron drift instabilities (comparable to Kelvin-Helmholtz instabilities).\cite{kent} Finally, collisions can also destabilize some particular modes, such as the lower hybrid instability.\cite{smolyakov}

In this Letter, a one-dimensional (1D) model of the cross field equilibrium plasma transport is developed. The model is linearized including time-dependence in order to describe the linear growth of the instability. An electrostatic particle-in-cell / Monte Carlo collision (PIC / MCC) code \cite{birdsall_book} LPPic \cite{croes_modelisation_2017, croes_psst,croes_iepc,lucken_psst,lucken_iepc} is used to simulate a magnetized plasma column in two-dimensional (2D) Cartesian coordinates. The spectral analysis of the PIC simulation results is compared with the solution of the theoretical dispersion relation, and the effective collision frequency measured in the PIC simulation is compared with the analytical model. 

The results are of importance for the modeling of industrial plasma devices such as neutral beam injectors,\cite{hagelaar} and electric space propulsion systems.\cite{thomas} \\

In this Letter, the frequencies are normalized to the electron cyclotron frequency $\omega_{ce} = e B / m_e$, where $B$ is a uniform magnetic field, the length dimensions are normalized to the Larmor radius of thermal electrons $\rho_L = \frac{1}{eB}(k_B T_e m_e)^{1/2}$, where $T_e$ is the electron temperature, that is assumed uniform, the velocities are normalized to the electron thermal velocity $v_{Te} = (k_B T_e/m_e)^{1/2}$, and the electrostatic potential $\phi$ is normalized to the electron temperature in volts. Under these conditions, the electron momentum balance equation is:
\begin{equation}
\frac{d\ve{v}}{dt} = \nabla\phi - \ve{v}\times\ve{b} - \frac{\nabla n}{n} - \nu \ve{v} 
\label{mom_elec0}
\end{equation}
where $d/dt$ is the Lagrangian time derivative, $\ve{v}$ is the electron fluid velocity, $n$ is the electron density, $\nu$ is the electron collision frequency, and $\ve{b}$ is the unit vector in the direction of the magnetic field.
It is assumed that the steady-state plasma properties depend only on the variable $x$ and that the magnetic field is along $z$ ($\ve{b} = \ve{e}_z$). We investigate a symmetrical single-ion plasma where the plasma density drops monotonically from the discharge center at $x=0$ to a floating sheath region, near $x = l/2$. 
The ions have a mass $m_i$, their temperature is much lower than the electron temperature, such that ion diffusion is neglected \correc{compared to the mobility driven flux}, and the plasma is quasineutral. 

At steady-state, and making the common assumption that electron inertia is negligible due to the small electron mass \cite{chabert_book,lieberman,allen,riemann, sternberg, fruchtman}, \corr{\eq{mom_elec0}} is 
\begin{align}
v_y &= -n'/n + \phi' - \nu v_x \label{momx}\\
v_y &= v_x / \nu               \label{momy}
\end{align}
\correc{ In \eq{momx}, the first term of the right-hand side represents the diamagnetic drift, the second term is the $E\times B$ drift, and the third term is the drift associated with collisions. } 

The \correc{normalized} electron continuity equation is 
\begin{equation}
\partial n / \partial t + \nabla\cdot(n\ve{v}) = n\nu_{iz}
\label{continu}
\end{equation}
where $\nu_{iz}$ is a \correc{normalized} electron impact ionization frequency. The continuity equation is the same for ions and leads to equal electron and ion fluxes along $x$, \correc{ under the assumption that the fluid velocity is zero for both species at the discharge center.} This allows to use the same fluid velocity $v_x$ for both species. We assume that the ion Larmor radius is larger than the discharge dimensions so the ions are not magnetized. 
In 1D, \eq{continu} becomes
\begin{equation}
v_x' + \frac{n'}{n}v_x = \nu_{iz}.
\label{vx}
\end{equation}
Eliminating $v_y$ from \eqs{momx}{momy}, we obtain
\begin{equation}
-\phi' = -\frac{n'}{n} - v_x\left(\frac{1}{\nu} + \nu \right)
\label{Efield}
\end{equation}

\correc{ A very simple collisionless momentum conservation equation is used here for the ions:\footnote{Ion collisions could be included here, with a discussion on how to compute the ion momentum transfer collision frequency, \cite{godyak,lucken_psst} but with no influence on the strongly magnetized regime. } }
\begin{equation}
\frac{v_x^2}{2\mu} = -\phi \Rightarrow \frac{v_x v_x'}{\mu} = -\phi' .
\label{momion}
\end{equation}
where $\mu = m_e/ m_i$.
The electric field and the pressure gradient terms are eliminated using \eqs{momion}{vx} respectively, and inserted in \eq{Efield}, which yields
\begin{equation}
\left( 1 - \frac{v_x^2}{\mu} \right)v_x' = \nu_{iz} + \correc{ \alpha } v_x^2. 
\label{diff_vx}
\end{equation}
where $\alpha = 1/\nu + \nu$.
This first order differential equation is integrated to obtain
\begin{equation}
(\alpha \mu + \nu_{iz})\arctan\left( \frac{v_x \alpha^{\frac{1}{2}}}{\nu_{iz}^{\frac{1}{2}}} \right) - (\alpha\nu_{iz})^{\frac{1}{2}} v_x = \mu\alpha^{\frac{3}{2}} \nu_{iz}^{\frac{1}{2}} x
\label{vx_implicit}
\end{equation}
\correc{
The quasineutral assumption, that is valid in the plasma bulk, breaks at the sheath edge. The boundary condition for the quasineutral plasma is chosen to be the Bohm criterion \cite{riemann} 
\begin{equation}
v_x = \mu^{1/2}
\label{bohm}
\end{equation}
that needs to be satisfied at the sheath edge $x=l/2$, even for magnetized plasmas \cite{allen, yankun}.
Equation~\ref{vx_implicit} yields the equation for the electron temperature
\begin{equation}
\left[ \left( \frac{\alpha \mu}{\nu_{iz}}  \right)^{\frac{1}{2}}  + \left( \frac{\nu_{iz}}{\alpha \mu}  \right)^{\frac{1}{2}}  \right]\arctan\left[ \left( \frac{\alpha \mu}{\nu_{iz}}  \right)^{\frac{1}{2}} \right] - 1 = \frac{\mu^{\frac{1}{2}}\alpha l}{2}
\label{electemp}
\end{equation}
that can easily be solved numerically for $\nu_{iz}(T_e)$. 
From \eq{momion}, the electric field is derived:
\begin{equation}
-\phi' = \frac{v_x\left( \nu_{iz} + \alpha v_x^2 \right) }{\mu - v_x^2 }
\label{elec_vx}
\end{equation}
and so is the plasma density profile from \eq{vx}
\begin{equation}
\frac{n}{n_0} = \left( 1 + \frac{\alpha v_x^2}{\nu_{iz}} \right)^{-\frac{1}{2}\left( 1 + \frac{\nu_{iz}}{\alpha \mu} \right) }
\label{normalizedn}
\end{equation}
}
\correc{ We now consider the high magnetic field limit in which $\alpha, x, l \gg 1$, for $x/l\neq 0$. 
\eq{electemp} yields
\begin{equation}
\nu_{iz} =  \frac{\pi^2}{\alpha l^2} 
\label{nuiz_HB}
\end{equation}
such that \eq{vx_implicit} is also }
\correc{
\begin{equation}
v_x = \left( \frac{\nu_{iz}}{\alpha}\right)^{\frac{1}{2}} \tan\left( \nu_{iz}^{\frac{1}{2}}\alpha^{\frac{1}{2}} x \right) \approx \frac{\pi}{\alpha l} \tan\left( \frac{\pi x}{l} \right) 
\label{vx_HB}
\end{equation}
At a given relative position $x/l < 1/2$, $v_x \underset{\alpha \rightarrow \infty}{\longrightarrow} 0$.  }\\

\correc{
The ratio between the electric field and the velocity term in \eq{Efield} is hence
\begin{equation}
\frac{\pi^2 / (\alpha l)^2 + v_x^2 }{\mu - v_x^2} \underset{\alpha \rightarrow \infty}{\longrightarrow} 0
\end{equation}
everywhere in the plasma bulk ($v_x < \mu^{1/2}$). 
This means that at high magnetic, the electron drift is dominated by the diamagnetic drift, except at the sheath edge, as will be shown below. 
}

\correc{
In the bulk, \eq{vx_HB} yields
\begin{equation}
v_x' =  \nu_{iz} / \cos^2\left( \nu_{iz}^{\frac{1}{2}}\alpha^{\frac{1}{2}} x \right) = \alpha v_x^2 / \sin^2\left( \nu_{iz}^{\frac{1}{2}}\alpha^{\frac{1}{2}} x \right)
\end{equation}
Using \eq{momion}, the electric field has a finite value at the sheath edge
\begin{equation}
-\phi' = \alpha \mu^{1/2}
\end{equation}
This is used in \eq{Efield} to estimate the diamagnetic drift term at the sheath edge
\begin{equation}
-n'/n = 2 \alpha \mu^{1/2}
\label{diamagnetic_HB}
\end{equation}
Hence, for $\alpha > 1/(2\mu^{1/2})$, the diamagnetic drift is larger than the electron thermal velocity. The diamagnetic drift is a purely fluid drift term and should typically not be higher than the thermal velocity. This intuition was confirmed by the PIC simulation and gives a hint that $\alpha$ cannot be arbitrarily large due to the magnetic field. If this is the case, then there should be an effective electron momentum transfer collision frequency $\nu_B$, that depends on the magnetic field, and that should satisfy roughly (in normalized dimensions) $\nu_B > 2\mu^{1/2}$, such that with $\alpha=1/\nu_B$, the right hand side of \eq{diamagnetic_HB} is less than unity. In general, this effective collision frequency may depend on $x$. However, it will be shown by the PIC simulation that the uniform effective collision model provides relevant corrections to the classical description. 
}

\correc{
In a 1D system, the $h$ factor that characterizes the ion losses to the walls is  $h = n_s /n_0$, where $n_s$ is the plasma density at the sheath edge. Integrating \eq{vx} leads to 
\begin{equation}
 \nu_{iz} = \pi h \mu^{1/2} / l 
\end{equation} 
where 
\begin{equation}
h = \frac{\pi }{ \alpha l \mu^{1/2}}
\end{equation}
because of \eq{normalizedn}. At high magnetic field,
\begin{equation}
\nu_{B} = h\mu^{1/2}l / \pi .
\label{nuB}
\end{equation}
In dimensional units,
\begin{equation}
\nu_{B} = \frac{h m_e \omega_{ce}^2 L}{\pi (m_i k_B T_e)^{1/2}}.
\label{nuB_denorm}
\end{equation}
}
where $L = \rho_L l$ is the denormalized system size. \corr{At this point, both the $h$ factor and the effective collision frequency may depend on the magnetic field, which is consistent with instability enhanced transport regimes.} \\

We now examine the plasma unstable behavior. It is assumed that a steady-state solution is perturbed by a harmonic wave propagating in the $y$ direction such that the first order densities $n_{e1}$ and $n_{i1}$ for electrons and ions respectively, and the first order potential $\phi_1$ are proportional to $\exp(-i\omega t + iky)$, where $\omega$ is the complex wave frequency (normalized to $\omega_{ce}$), and $k$ is the wavenumber (normalized to $\rho_L^{-1}$).
\correc{ Collisions were added to the classical inhomogeneous fluid plasma theory\cite{mikhailovskii} (see Appendix), and} the perturbed electron density response to the potential is given by \correc{the susceptibility}
\begin{equation}
\chi_e \equiv \frac{\omega_p^2}{k^2}\frac{n_{e1}}{\phi_1} = \frac{\omega_p^2}{k^2}\frac{\omega_* + k^2(\omega + \omega_0 + i\nu)}{(\omega + \omega_0) + k^2(\omega + \omega_0 + i\nu)}
\label{chi_e}
\end{equation}
where
\begin{equation}
 \omega_0 = -k \phi'  \hspace{5mm} , \hspace{5mm} \omega_* = -k n'/n
 \label{w0s_def}
\end{equation}
and $\omega_p^2 = n_0 e^2 / (\epsilon_0 m_e \omega_{ce}^2)$ is the normalized electron plasma frequency.
Equation~(\ref{chi_e}) is also introduced by the gyroviscosity formalism in Smolyakov et al..\cite{smolyakov} The cold, collisionless, non-magnetized ions do not drift in the $y$ direction. Their susceptibility is \cite{bellan}
\begin{equation}
\chi_i \equiv -\frac{\omega_p^2}{k^2}\frac{n_{i1}}{\phi_1} = -\frac{\mu \omega_p^2}{\omega^2}.
\label{chi_i}
\end{equation}
The first order Poisson's equation \cite{bellan}
\begin{equation}
1 + \chi_e + \chi_i = 0
\label{disp_chi}
\end{equation}
provides the wave dispersion relation. Introducing the polynomials
\begin{align}
P_0(\omega) = \, & (\omega+\omega_0)\left[ (1+k^2+\omega_p^2)\omega^2 - \mu\omega_p^2(1+k^2) \right] \nonumber \\
                 & + \omega^2\omega_p^2\omega_* / k^2                                                          \\
Q(\omega)   = \, & (\omega_p^2 + k^2)\omega^2 - \mu\omega_p^2k^2,
\end{align}
the dispersion relation is
\begin{equation}
P(\omega) = P_0(\omega) + i \nu Q(\omega) = 0.
\label{disp_poly}
\end{equation}
Equation~(\ref{disp_poly}) is valid as long as the frequency of the instability is smaller than the electron cyclotron harmonics, where a kinetic description of the Bernstein modes is required.\cite{allis, schmidt, bellan,lafleur_II} Inspecting the following table, $P_0$ has always two negative roots and one positive root: 
\begin{center}
\begin{tabular}{c | c| c| c| c}
$\omega$      & $-\infty$ & $-\omega_0$ & 0    & $+\infty$  \\
$P_0(\omega)$ & $<0$      & $>0$        & $<0$ & $>0$  
\end{tabular}
\end{center}
Let $\omega_r$ be the positive root. Since $\nu\ll 1$, collisions can be treated as a perturbation term.
The perturbed solution being $\omega_r + i\nu\omega_i$, to the first order in $\nu$, 
\begin{equation}
P(\omega_r + i\nu\omega_i) = 0 \Leftrightarrow \omega_i = - Q(\omega_r) / P_0'(\omega_r)
\end{equation}
Since $P_0'(\omega_r)>0$, the mode stability is determined by the sign of $Q(\omega_r)$.
If $\omega_* = \omega_0 = 0$, the positive root is 
\begin{equation}
\omega_r = \mu^{1/2}\left( \frac{1}{\omega_p^2} + \frac{1}{1+k^2} \right)^{-1/2}   ,
\label{nodrift}
\end{equation}
 a mode transiting from the lower hybrid frequency at low $k$'s to the ion plasma frequency at high $k$'s, and damped by collisions. 
If $\omega_0 = 0$, but $\omega_* \neq 0$, the solution is 
\begin{equation}
\omega_r = \frac{\omega_*\omega_p^2 \left\{ \left[ 1 + \frac{4\mu k^4(1+k^2)(1+k^2+\omega_p^2)}{\omega_p^2\omega_*^2} \right]^{\frac{1}{2}} -1 \right\}}{2 k^2(1+k^2 + \omega_p^2)} 
\label{noExB}
\end{equation}
The general instability criterion is found by solving jointly $P_0(\omega) = 0$ (the mode exists) with $Q(\omega)=0$ (the mode is at stability limit):
\begin{equation}
\omega_* - \omega_0 = \frac{\mu^{1/2}k}{(1 + k^2/\omega_p^2)^{1/2}}.
\label{instab_crit}
\end{equation}
This means that the plasma is unstable when the fluid electron drift is higher than the ion sound speed. The numerical resolution of the two other roots of $P$ shows that they are all stable in the regime of interest here. 
Using \eqs{w0s_def}{Efield} for low wavenumbers, the plasma is unstable if:
\begin{equation}
v_x \left(1/\nu + \nu \right) > \mu^{1/2}  .
\end{equation}

At the sheath edge, $v_x = \mu^{1/2}$, and $1/\nu + \nu$ is greater than 1 for all $\nu>0$. Therefore, the plasma is always unstable, at least at the sheath edge, as long as the electrons are magnetized ($l \gg 1$). The destabilization of similar modes by collisions was first found experimentally \cite{birdsall_resistive, hendel} and explained theoretically by Chen \cite{chen_universal} as a particular type of resistive drift mode.\cite{chen_instab} More recently, several experiments highlighted drift wave instabilities in magnetized plasma columns.\cite{brochard, gravier} \correc{ As shown by Lakhin et al. taking into account finite Larmor radius effects,\cite{lakhin_effects_2018,lakhin_marginal_2018} the condition $-n'\phi'/n < \mu / 4$ is sufficient to ensure electrostatic modes stability in a collisionless plasma, and only collisions can destabilize it in this case. }

In order to validate the 1D model, the PIC/MCC simulation needs to be 2D to capture both the direction of the instability and the gradients of the equilibrium solution. Simulations of a square argon discharge were performed at gas pressures of 3, 6, and 12\,mTorr and uniform magnetic fields between 0 and 40\,mT oriented along $z$, perpendicularly to the simulation plane. 
The discharge is sustained by a uniform radio-frequency (RF) electric field at 13.56\,MHz along $z$. The amplitude of the heating electric field is adjusted at each RF cycle to control the power absorbed by the plasma, fixed to 9.6\,kW/m$^3$. The $x$ and $y$ components of the electric field come from the solution of the Poisson's equation, with the potential set to zero at the conducting walls. 

Super-particles lost at the walls are discarded and new particles are generated through self-consistent ionization with a uniform background gas at 300\,K. The newly created ions and electrons are initialized with Maxwellian distribution functions at 0.026\,eV and 4\,eV respectively. The ion temperature remains lower than 0.2\,eV in all simulation conditions. The steady-state electron temperature is nearly uniform in the bulk plasma and varies between 3.6 and 5.3\,eV depending on the cases. 
A realistic kinetic scheme for argon is used \cite{lucken_iepc} using cross sections from the LXCat database.\cite{biagi, phelps, phelps_application_1994} \corr{The simulation time step and the cell size satisfy the classical stability criteria (Birdsall and Langdon \cite{birdsall_book}). The number of particles per cell was always greater than 100 in the center (150 in most cases). A simulation at 20\,mT and 3\,mTorr was performed with 4 times more particles per cell (600 part./cell in the discharge center) with a change of less than 3\% for the discharge global properties. }

Plasma sheaths form in about one microsecond, and at high magnetic field, instabilities rotating in the electron drift direction develop rapidly. After a transient of about 20\,\textmu s (depending on the pressure and the magnetic field), the volume-averaged plasma properties reach a steady-state and the instability saturates with a continuous spectrum featuring a maximum around 3\,MHz. The instability wavevector is directed mainly in the azimuthal direction.

\begin{figure}
\includegraphics[width=\linewidth]{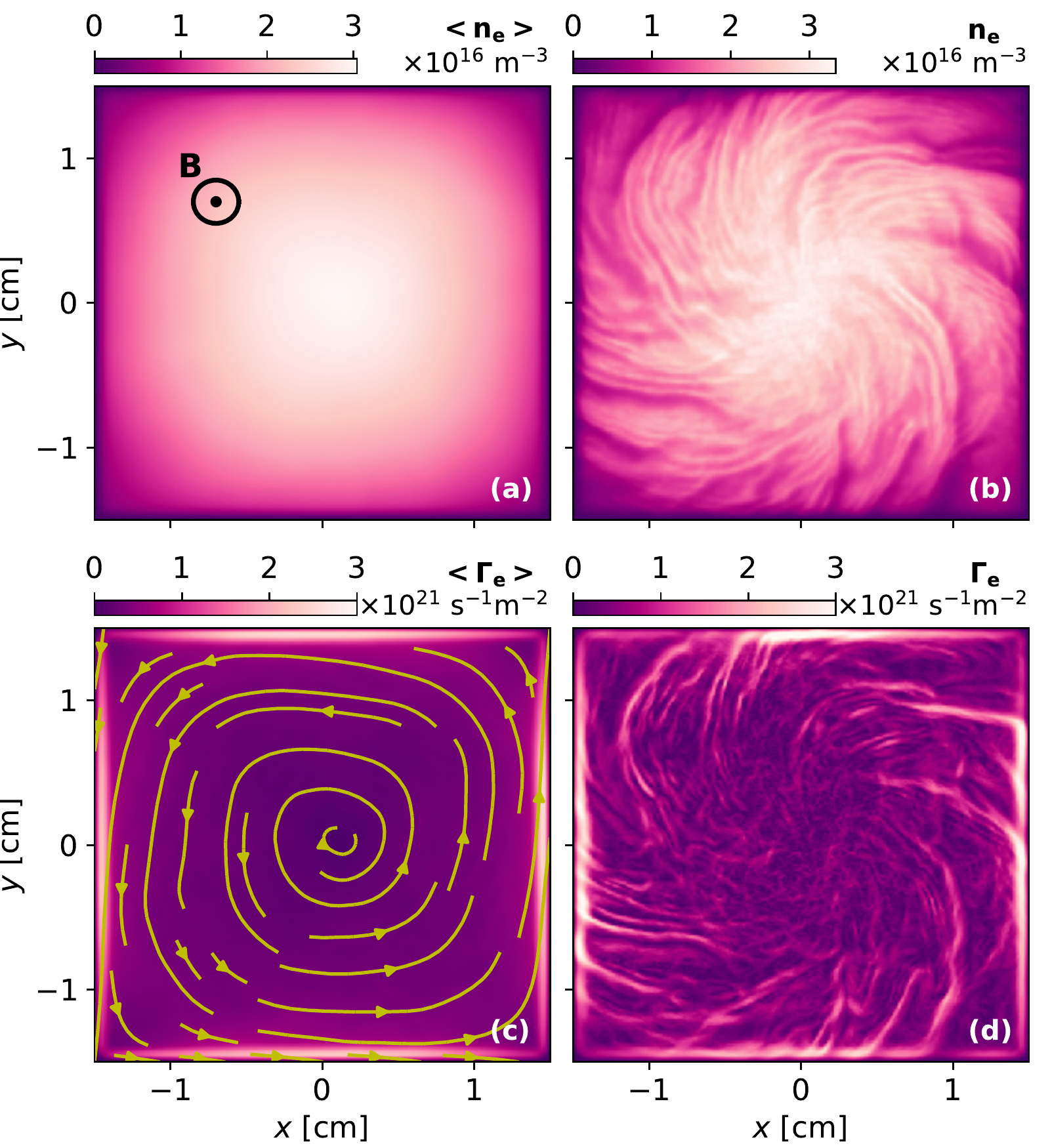}
\caption{\label{fig:xyplots} Electron density and norm of the electron flux at 91\,\textmu s (b, d), and averaged over the last 27\,\textmu s of the simulation (a, c). In (c), the spiral streamlines represent the electron flux. The data come from a 3\,mTorr, 20\,mT LPPic simulation.} 
\end{figure}

\fig{fig:xyplots} shows the electron density and the electron current at 3\,mTorr and 20\,mT. When the plasma properties are averaged over several tens of microseconds, the discharge aspect is the same as in the non-magnetized case,\cite{lucken_psst} the only qualitative difference being that the electrons are rotating in the direction of the diamagnetic drift. On a shorter time-scale, the instabilities are visible and the waves seem to break against the sheath. 

\begin{figure}
\vspace{2mm}\includegraphics[width=\linewidth]{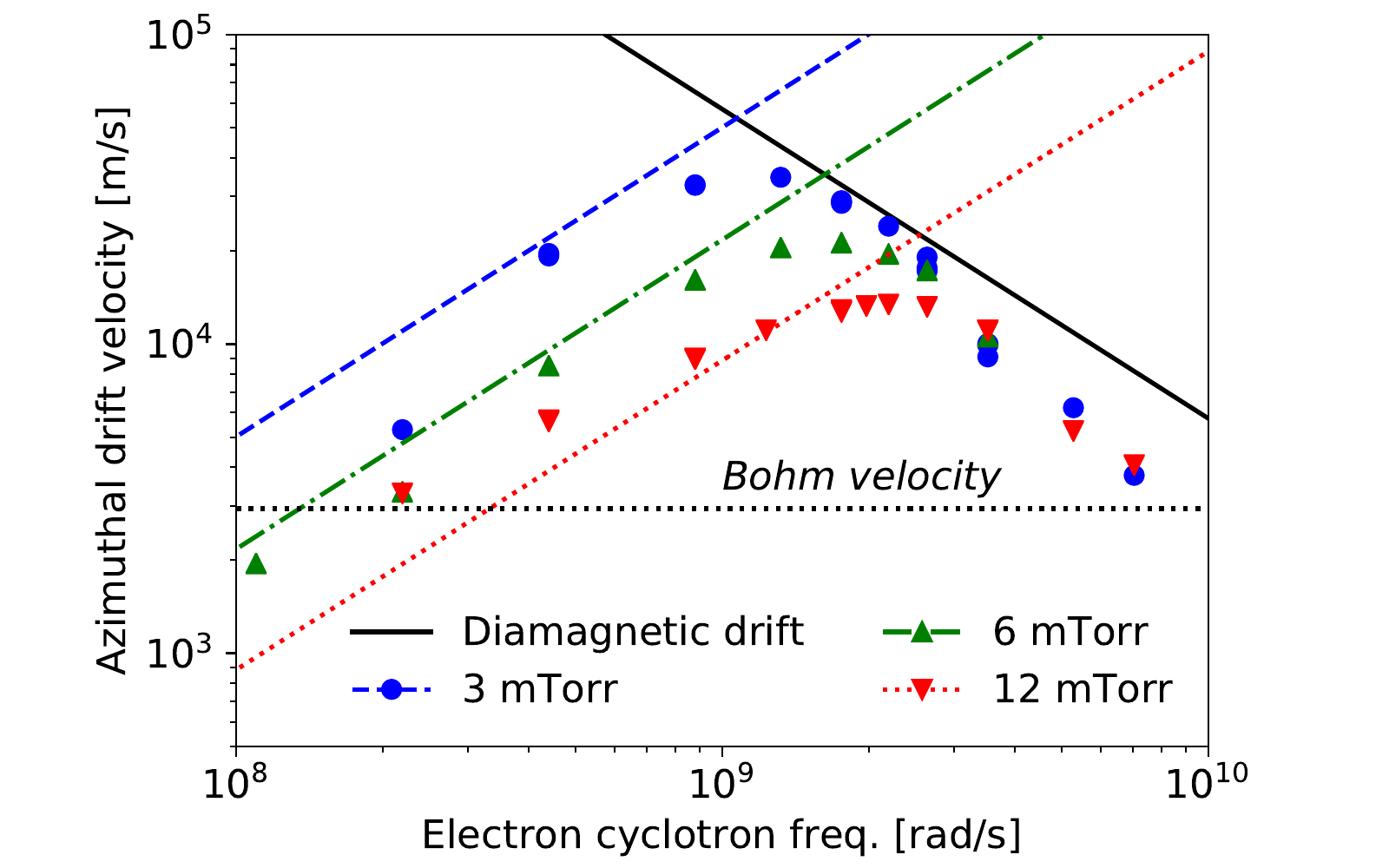}
\caption{\label{fig:drifts} Azimuthal drift velocities measured for various values of the pressure and magnetic field, measured at $r = 9$\,mm from the discharge center. The various dashed lines correspond to the classical regime (\eq{vth_class}), and the solid black line corresponds to the instability dominated regime described by \eq{vth_instab}. }
\end{figure}
In \fig{fig:drifts}, the electron azimuthal drifts measured in the PIC simulations are compared with simple asymptotic formulae coming from the model equations. Neglecting curvature effects, the azimuthal direction is treated as $y$, and the radial direction as $x$.
The colored dashed lines in \fig{fig:drifts} correspond to the weakly-magnetized limit of \eq{vx_HB}, 
\begin{equation}
v_{\theta} = \frac{h\mu^{1/2}}{\nu}\tan\left( \frac{\pi r }{l} \right)
\label{vth_class}
\end{equation}
for each value of the pressure. The following formula coming from 2D theories of non-magnetized plasma transport was used for the $h$ factor \cite{lucken_psst}
\begin{equation}
h = 0.55 \left[3+0.5\frac{L}{\lambda_i} + 0.2 \frac{T_i}{T_e}\left(  \frac{L}{\lambda_i}\right)^2\right]^{-1/2}.
\label{h0} 
\end{equation}
where $\lambda_i$ is the (dimensional) ion mean free path and $T_i$ is the ion temperature.
The solid black line in \fig{fig:drifts} corresponds to the diamagnetic drift
\begin{equation}
v_{\theta} = \frac{\pi}{l} \tan\left( \frac{\pi r }{l} \right)
\label{vth_instab}
\end{equation}
For simplicity, a reference temperature of 3.6\,eV was considered in all cases. 

\begin{figure}
\includegraphics[width=\linewidth]{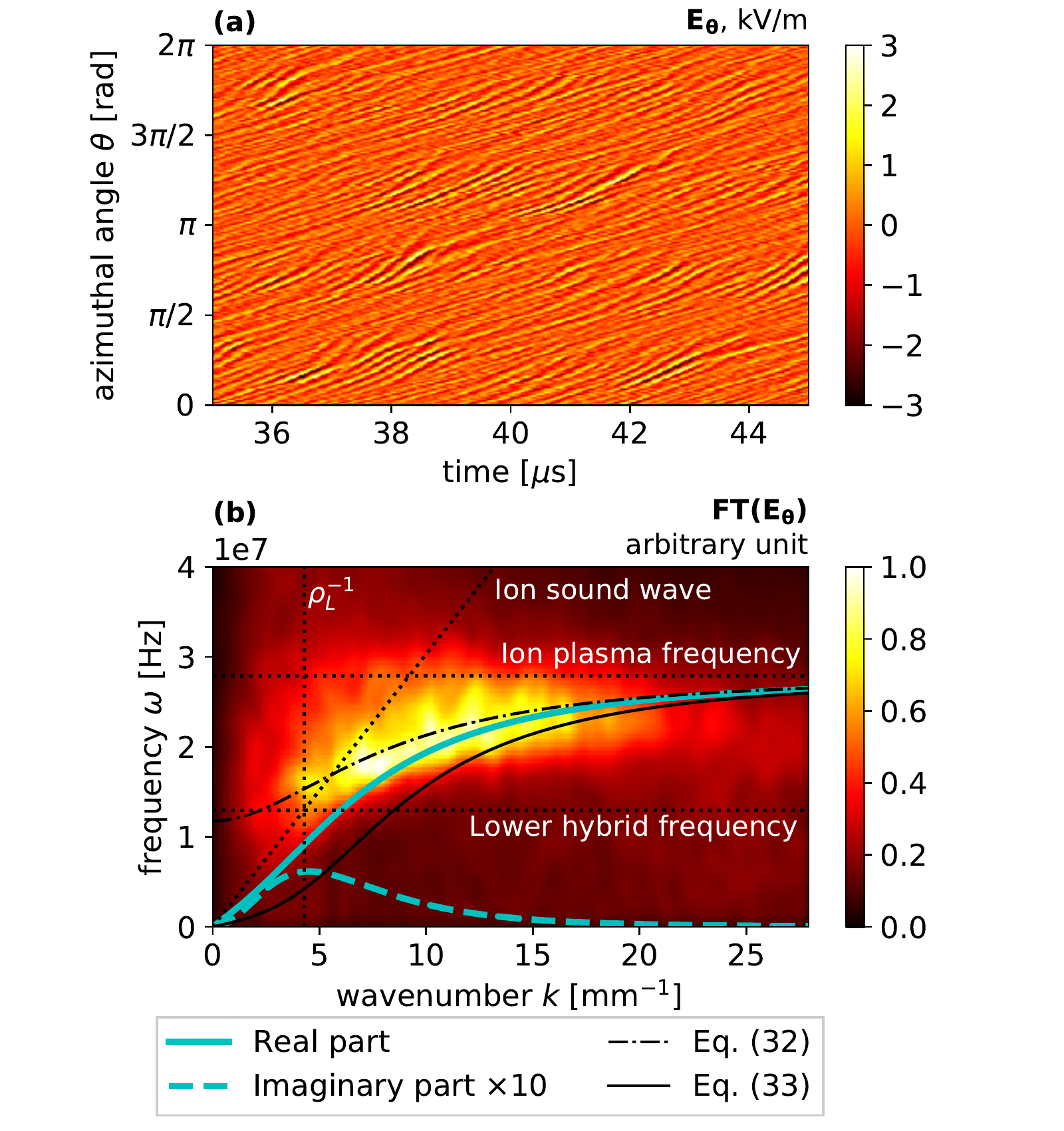}
\caption{\label{fig:Eth} (a) Azimuthal electric field obtained at 3\,mTorr, and 20\,mT, at a distance of 9\,mm from the center of the simulation domain. (b) The corresponding spatio-temporal FT (colorplot) with a numerical solution of Eq.~(\ref{disp_poly}) in cyan, approximate solutions in dashed and solid black lines, with the parameters of \tab{tabinstab}.}
\end{figure}

\begin{table}
\begin{tabular}{l c r}
\toprule 
Magnetic field       & 20                   & mT       \\
Pressure             & 3                    & mTorr    \\
Plasma density       & $1.8\times 10^{16}$  & m$^{-3}$ \\
Electron temperature & 3.84                 & eV       \\
Diamagnetic drift    & 17.3                 & km/s     \\  
$E\times B$ drift    & -4.3                 & km/s     \\
\toprule
\end{tabular}
\caption{\label{tabinstab} Numerical quantities of the simulation used to solve the dispersion relation.}
\end{table}

\fig{fig:Eth}(a) shows the azimuthal electric field as a function of time and $\theta$ in a typical case where the instability drives the transport. The 2D Fourier transform (FT) of this signal is shown in \fig{fig:Eth}(b). The dashed and solid black lines are the approximate solutions of \eqs{nodrift}{noExB} respectively, while the full numerical solution of \eq{disp_poly} using the parameters of \tab{tabinstab} is displayed in cyan. The electron density, temperature, and drifts are extracted from the time-averaged solution of the PIC simulation in saturated state. 
\begin{figure}
\includegraphics[width=\linewidth]{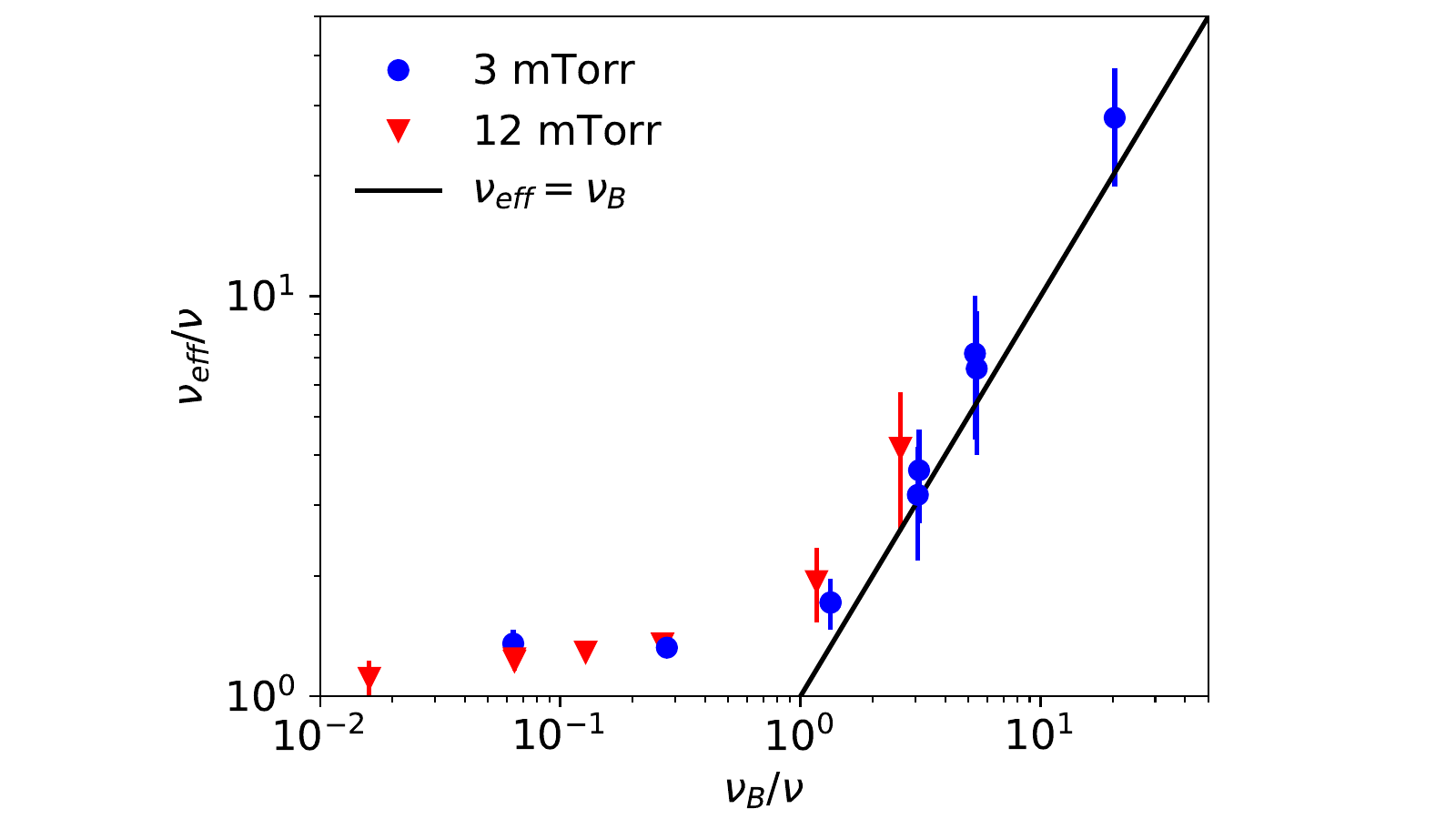}
\caption{\label{fig:nueff} Effective electron collision frequency (\eq{nueff}) plotted against the instability-enhanced collision frequency predicted by \eq{nuB} \corr{using \eq{h0}}. }
\end{figure}

An effective collision frequency is extracted from the simulation data using \eq{momy}
\begin{equation}
\nu_{\rm eff} = \langle \Gamma_{er} \rangle / \langle \Gamma_{e\theta} \rangle
\label{nueff}
\end{equation}
where $\langle \Gamma_{er} \rangle$ and $\langle \Gamma_{e\theta} \rangle$ are the electron fluxes in the radial and azimuthal directions respectively, averaged along $\theta$. The effective collision frequency of \eq{nueff} is normalized to the classical collision frequency known from the literature \cite{lieberman} and plotted in \fig{fig:nueff}. The error bars correspond to all the data collected at various radial positions, from 1 to 14\,mm from the discharge center. 
At low magnetic field, the effective collision frequency is equal to the classical collision frequency. In the strongly magnetized regime, the effective collision frequency is correctly \corr{described} by \eq{nuB_denorm}, \corr{assuming an $h$ factor that does not depend on the magnetic field} \correc{(\eq{h0})}. \\

A 1D model of the cross field plasma transport was derived and validated using 2D PIC simulations. \correc{ It was shown that if the diamagnetic drift at the sheath edge remains below the electron thermal velocity}, then the effective electron collision frequency should depend on the magnetic field. This result is consistent with the plasma unstable behavior predicted by the linear theory of wave perturbations, and by PIC simulations. The effective collision frequency scales as $\omega_{ce}^2$ at high magnetic field. 
\correc{ The transport across a magnetic field in a weakly ionized plasma is much better described by assuming that the $h$ factor does not depend on the magnetic field than by assuming a classical collision frequency}.\footnote{The dependence of the $h$ factor on the magnetic field, from the classical regime to the instability-enhanced regime, will be the focus of another paper.  }$^,$\cite{sternberg}
The plasma transport becomes completely insensitive to the magnetic field due to the electron drift resistive instability. \\

{\it Acknowledgments -- }
The authors are grateful to A.\,J. Lichtenberg, A. Tavant, R. Martorelli, and O. Gurcan for many enlightening discussions. This work was granted access to the HPC resources of CINES under the allocation 2017-A0020510092 made by GENCI, and it was partially funded by the Agence Nationale de la Recherche (ANR-16-CHIN-0003-01) and the U.S. Department of Energy Office of Fusion Science Contract DE-SC0001939. \\

{\it Appendix: Electron response to a perturbed electrostatic potential \\}
The normalized electron momentum conservation equation is
\begin{equation}
\frac{d\ve{v}}{dt} = \nabla\phi - \ve{v}\times\ve{b} - \frac{\nabla n}{n} - \nu \ve{v}.
\label{mom_elec}
\end{equation}
Assuming
\begin{align}
\nu, d/dt = \mathcal{O}(\epsilon)   \label{ordering1} \\
\phi, n, |\nabla| = \mathcal{O}(1)  \label{ordering0}
\end{align}
where $\epsilon$ is a small parameter, we can expand $\ve{v}$ with respect to $\epsilon$.
To the 0-th order, 
\begin{equation}
\nabla\phi - \ve{v} \times \ve{b} - \nabla n / n = 0
\end{equation}
which yields the $E\times B $ and diamagnetic drifts, respectively 
\begin{equation}
\ve{v}_E = \ve{b}\times \nabla\phi
\text{ ~~ and ~~  }
\ve{v}_d = \frac{\nabla n }{n}\times \ve{b}.
\end{equation}
\eq{mom_elec} is then to the first order
\begin{equation}
\frac{d}{dt}(\ve{v}_E + \ve{v}_d) = -\ve{v}^{(1)}\times \ve{b} - \nu (\ve{v}_E + \ve{v}_d)
\end{equation}
where $\ve{v}^{(1)} = \ve{v} - \ve{v}_E - \ve{v}_d$ is the first order term of the electron velocity.
Hence,
\begin{equation}
\ve{v}^{(1)} = \ve{v}_p + \ve{v}_{dp}
\end{equation}
where 
\begin{equation}
\ve{v}_p = \left( \frac{d}{dt} + \nu \right)\ve{v}_E \times \ve{b} = \left( \frac{d}{dt} + \nu \right)\nabla \phi
\end{equation}
and 
\begin{equation}
\ve{v}_{dp} = \left( \frac{d}{dt} + \nu \right)\ve{v}_d \times \ve{b} = - \left( \frac{d}{dt} + \nu \right) \frac{\nabla n }{n}
\end{equation}
are the polarization drift terms due to the $E\times B$ and the diamagnetic drifts, respectively (including the friction force).
It is useful to derive the fluxes corresponding to each of the drift terms.
\begin{equation}
\nabla\cdot(n\ve{v}_E) = (\ve{b}\times \nabla \phi)\cdot\nabla n + n \nabla\cdot (\ve{b}\times \nabla \phi)
\end{equation}
The second term is zero due to the formula 
\begin{equation}
\nabla\cdot(u\times v) = -u\cdot(\nabla\times v) + v\cdot(\nabla\times u).
\label{div_cross}
\end{equation} 
Thus, 
\begin{equation}
\nabla\cdot(n\ve{v}_E) = (\ve{b}\times \nabla \phi)\cdot\nabla n
\label{ExB_flux}
\end{equation}
Using again \eq{div_cross}, 
\begin{equation}
\nabla\cdot(n\ve{v}_d) = 0.
\end{equation}
The relationship is known as the gyroviscous cancellation.\corr{\cite{smolyakov}}
By neglecting all particle source and loss terms ($\nu_{iz} \ll \omega$), the electron continuity equation is
\begin{equation}
\frac{\partial n}{\partial t } + \nabla \cdot (n\ve{v}) = 0.
\label{elec_cont}
\end{equation}
So to the first order in $\epsilon$,
\begin{equation}
\frac{\partial n}{\partial t } + \nabla\cdot[n(\ve{v}_E + \ve{v}_d + \ve{v}_{p} + \ve{v}_{dp})] = 0,
\end{equation}
or
\begin{equation}
\frac{\partial n }{\partial t} + (\ve{b}\times \nabla \phi)\cdot\nabla n + \nabla \cdot \left[ n \left( \frac{d}{dt} + \nu \right) \left( \nabla\phi - \frac{\nabla n}{n} \right) \right] = 0
\label{tot_elec}
\end{equation}

The only term contributing to the motion of the electron guiding centers is the $E\times B$ drift \cite{weiland}
\begin{equation}
d/dt = \partial / \partial t + \ve{v}_E \cdot \nabla 
\end{equation}
Using the property
\begin{equation}
\nabla\cdot[(\ve{v}_E\cdot\nabla)\nabla\phi] = (\ve{v}_E\cdot\nabla)\nabla^2 \phi,
\end{equation}
\begin{align}
\nabla\cdot(n\ve{v}_p) =\, & n\left(\frac{\partial }{\partial t} + \ve{v}_E \cdot \nabla + \nu \right) \nabla^2\phi   \nonumber   \\
                         & + \nabla n\cdot \left(\frac{\partial }{\partial t} + \ve{v}_E \cdot \nabla + \nu \right) \nabla\phi.
\end{align}
For the density gradient polarization drift,
\begin{align}
\nabla\cdot(n\ve{v}_{dp}) =& \nabla\cdot\left[ \frac{\nabla n}{n}\left( \frac{\partial}{\partial t} + \ve{v}_E\cdot \nabla \right) n \right]  \\
                           &- \left(\frac{\partial}{\partial t } + \nu\right)\nabla^2 n - \nabla\cdot[(\ve{v}_E\cdot\nabla)\nabla n] . \nonumber
\end{align}
We assume 
\begin{equation}
n = n_0(x)    + n_1(y, t) \hspace{3mm} ; \hspace{3mm} \phi = \phi_0(x) + \phi_1(y, t)
\end{equation}
with $n_1 \ll n_0$ and $\phi_1 \ll \phi_0$, and $n_1$ and $\phi_1$ proportional to $\exp(-i\omega t + ik y)$.
To the first order in $\phi_1$ and $n_1$: 
\begin{align}
\nabla\cdot(n\ve{v}_p) =& \,  n_0\left[ \frac{\partial}{\partial t} + ( \ve{b}\times\nabla\phi_0 \cdot\nabla ) + \nu \right]\nabla^2\phi_1  \nonumber \\
                        &+ \nu n_1 \nabla^2\phi_0 + n_0 ( \ve{b}\times\nabla\phi_1 \cdot\nabla ) \nabla^2\phi_0                              \nonumber \\
                        &+ \nabla n_0 \cdot ( \ve{b}\times\nabla\phi_1 \cdot\nabla ) \nabla\phi_0                             \\
                       =& \, i n_0 k^2(\omega + \omega_0 + i\nu)\phi_1 + \nu n_1 \phi_0''     \nonumber \\                      
                        &- ik n_0 \phi_0'''\phi_1 + i\omega_*n_0 \phi_0''\phi_1
\end{align}
where $\omega_* = -k n_0'/n_0$ and $\omega_0 = -k \phi_0'$.
Similarly, 
\begin{align}
\nabla\cdot(n\ve{v}_{dp}) =& \, \nabla\cdot\left\{ \frac{\nabla n_0 }{n_0} \left[ \frac{\partial }{\partial t} + (\ve{b}\times\nabla\phi_0\cdot\nabla)  \right]n_1 \right\} \nonumber \\
                           &+ \nabla\cdot\left[ \frac{\nabla n_0}{n_0}(\ve{b}\times\nabla\phi_1\cdot\nabla ) n_0 \nonumber \right] \nonumber \\
                           &- \left(\frac{\partial}{\partial t} + \nu\right)\nabla^2 n_1  
                           - \nabla\cdot\left[ (\ve{b}\times\nabla\phi_0\cdot\nabla)\nabla n_1 \right] \nonumber \\
                           &- \nabla\cdot\left[ (\ve{b}\times\nabla\phi_1\cdot\nabla)\nabla n_0 \right]  \\
                          =& -i n_1(\omega + \omega_0)\left(\frac{n_0''}{n_0} - \frac{\omega_*^2}{k^2} \right) - in_1\omega_*\phi_0'' \nonumber \\
                           &+ i \omega_*n_0\phi_1 \left( 2\frac{n_0''}{n_0} - \frac{\omega_*^2}{k^2}\right) \nonumber \\
                           &-i k^2(\omega + \omega_0 + i\nu)n_1 + ik\phi_1n_0''' .
\end{align}
Finally, \eq{ExB_flux} yields
\begin{equation}
\nabla\cdot(n\ve{v}_E) = -i\omega_0 n_1 + i\omega_* n_0 \phi_1
\end{equation}
To the first order in $\omega_*$, and neglecting second and third order derivatives of $n_0$ and $\phi_0$ (no shear), 
\begin{align}
\nabla\cdot(n \ve{v}_{p})  &=  i n_0 k^2(\omega + \omega_0 + i\nu)\phi_1 \\
\nabla\cdot(n \ve{v}_{dp}) &= -i k^2(\omega + \omega_0 + i\nu)n_1 
\end{align}
\eq{elec_cont} is therefore:
\begin{align}
&-i\omega n_1 - i\omega_0 n_1 + i\omega_* n_0 \phi_1 + i n_0 k^2(\omega + \omega_0 + i\nu)\phi_1  \nonumber \\
&- i k^2(\omega + \omega_0 + i\nu)n_1 = 0 
\end{align}

Hence,
\begin{equation}
\frac{n_1}{n_0} = \frac{\omega_* + k^2(\omega + \omega_0 + i\nu)}{\omega + \omega_0 + k^2(\omega + \omega_0 + i\nu)}\phi_1  .
\end{equation}
\corr{This result is valid within the assumptions of the model derived here (\eqs{ordering1}{ordering0}), but can be generalized by the Pad\'e approximation \cite{pegoraro,schep} to all wavenumbers. }

\bibliographystyle{apsrev4-1.bst}
\bibliography{biblio}

\end{document}